\documentclass{PoS}

\usepackage{xspace}
\usepackage{amsmath}
\usepackage{lineno}

\def \electron {$e^-$\xspace}
\def \positron {$e^+$\xspace}
\def \eeflux   {$e^-+e^+$\xspace}

\title{Towards a Measurement of the \positron\electron Flux above~1~TeV with
HAWC}

\ShortTitle{HAWC \positron\electron Flux}

\author{\speaker{Segev BenZvi}$^a$, Daniel Fiorino$^b$, Zigfried
Hampel-Arias$^b$, and Mehr Un Nisa$^a$, for the HAWC Collaboration$^c$\\
\llap{$^a$}Department of Physics and Astronomy, University of Rochester, Rochester, NY, USA \\
\llap{$^b$}Department of Physics and WIPAC, University of Wisconsin-Madison, Madison, WI, USA \\
\llap{$^c$}For a complete author list, see \href{http://www.hawc-observatory.org/collaboration/icrc2015.php}{www.hawc-observatory.org/collaboration/icrc2015.php}. \\
Email: \email{sybenzvi@pas.rochester.edu}, \email{fiorino@icecube.wisc.edu}, \email{zhampel@icecube.wisc.edu}, \email{mehr@pas.rochester.edu}}

\abstract{
  The High-Altitude Water Cherenkov (HAWC) Observatory records the air showers
  produced by cosmic rays and gamma rays at a rate of about 20~kHz. While the
  events observed by HAWC are 99.9\% hadronic cosmic rays, this background can
  be strongly suppressed using topological cuts that preferentially select
  electromagnetic air showers. Using this capability of HAWC, we can create a
  sample of air showers dominated by gamma rays and cosmic electrons and
  positrons. HAWC is one of the few operating observatories capable of
  measuring showers produced by \electron and \positron primaries above 1 TeV,
  and can record these showers from $2/3$ of the sky each day. We describe the
  sensitivity of HAWC to leptonic cosmic rays, and discuss prospects for the
  measurement of the \positron\electron flux and possible approaches for 
  \positron and \electron charge separation with the HAWC detector.}

\FullConference{The 34th International Cosmic Ray Conference,\\
		30 July- 6 August, 2015\\
		The Hague, The Netherlands}

\begin{document}

\section{Introduction}

\begin{figure}[ht]
  \includegraphics[width=\textwidth]{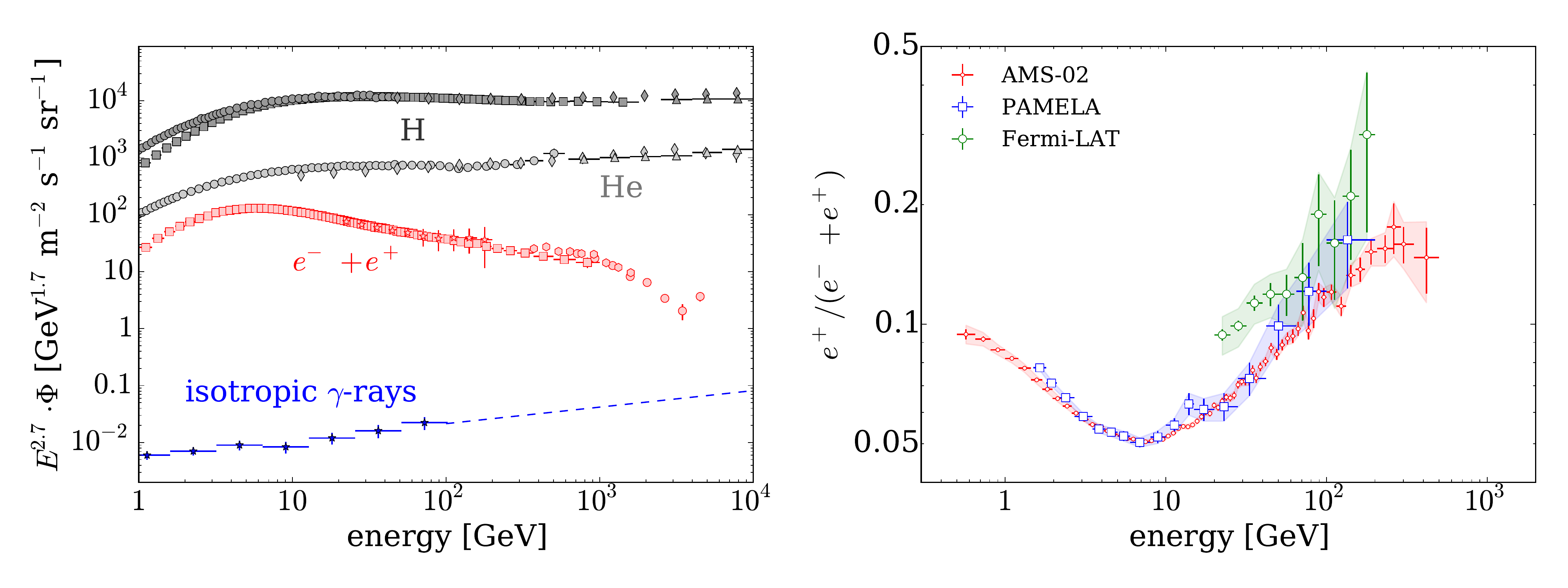}
  \caption{\label{fig:electron_positron_flux}
  {\sl Left}: the isotropic flux of cosmic protons
  \cite{Adriani:2011cu,Panov:2011ak,Yoon:2011aa,Aguilar:2015ooa}, helium
  \cite{Adriani:2011cu,Panov:2011ak,Yoon:2011aa}, and electrons and positrons
  \cite{FermiLAT:2011ab,Aguilar:2014fea,Aharonian:2008aa,Aharonian:2009ah} at
  GeV and TeV. The extragalactic isotropic diffuse flux of $\gamma$-rays
  \cite{Abdo:2010nz} and its extrapolation to TeV, another source of background
  for the $e^+e^-$ flux, are also plotted.
  {\sl Right}: the positron fraction measured from $0.5$ to $350$~GeV
  \cite{Adriani:2008zr,FermiLAT:2011ab,Aguilar:2013qda}. The error bars
  indicate statistical uncertainties, while the solid shaded regions show the
  total uncertainties (quadrature sum of statistical and systematic
  uncertainties).}
\end{figure}

The flux of cosmic electrons and positrons at Earth is a signficant component
of the all-particle flux below 10~GeV. However, due to synchrotron and inverse
Compton losses, this leptonic flux decreases faster than the $E^{-2.7}$ flux of
hadronic cosmic rays at higher energies.  Recent measurements of the combined
\eeflux flux from satellites \cite{FermiLAT:2011ab,Aguilar:2014fea} and
ground-based experiments \cite{Aharonian:2008aa,Aharonian:2009ah} indicate that
the spectrum decreases as $E^{-3.05}$ with a cutoff near $800$~GeV
(Fig.~\ref{fig:electron_positron_flux}). As a result, the \eeflux flux at 1~TeV
is approximately $0.1\%$ of the flux of hadronic particles.

While the rapidly falling \eeflux spectrum makes observations at TeV very
challenging, this energy range is of considerable astrophysical interest.  The
increase in radiative losses as a function of energy implies that above several
hundred GeV, the electrons and positrons observed at Earth must originate in
Galactic sources $<1$~kpc from the solar system \cite{Kobayashi:2003kp}. Hence,
the characterization of the spectrum at TeV is well-motivated, since features
in the spectrum can be used to study the closest accelerators of cosmic
electrons.

The relative abundance of positrons to electrons, usually expressed as the
ratio $e^+/(e^-+e^+)$, also contains important information about the origin of
the leptonic flux at Earth.  Measurements of \electron and \positron at
energies between $1$ to $100$~GeV range have shown that the fraction of
positrons in the flux increases from $5\%$ at $10$~GeV to $\sim15\%$ at
$100$~GeV \cite{FermiLAT:2011ab,Adriani:2008zr,Aguilar:2013qda}. This
observation runs counter to the expectation that cosmic antiparticles are
produced in secondary interactions \cite{Moskalenko:1997gh} and indicates that
the particles originate in a nearby source of primary cosmic rays. The origin
of the positron excess is not yet understood, with explanations ranging from a
local ($\sim100$~pc) source of primary \electron and \positron such as the
Geminga supernova remnant \cite{Yuksel:2008rf} to the production of $e^+e^-$ in
dark matter annihilation.

\section{The HAWC Observatory}

The High Altitude Water Cherenkov Observatory, or HAWC, is a gamma-ray and
cosmic-ray detector located $4100$~m above sea level in Sierra Negra, Mexico.
HAWC is an air shower array comprising 300 close-packed water Cherenkov
detectors (WCDs).  Each WCD is a light-tight steel tank containing $200$~kL of
purified water and four hemispherical photomultiplier tubes (PMTs).

The PMTs are used to detect the Cherenkov light produced when air shower
particles pass through the water in the tanks.  By combining the timing
information and spatial pattern of the PMTs triggered by an air shower, the
arrival direction and type of the primary particle can be identified.

For example, using simple topological cuts it is possible to discriminate air
showers produced by hadronic cosmic rays from the electromagnetic air showers
produced by gamma rays, electrons, and positrons.  The gamma-hadron
discrimination can be used to suppress the very large background of cosmic rays
($20$~kHz event rate) enough to produce sky maps of gamma-ray sources.  More
details on the operation of the HAWC detector, the event reconstruction, and
background suppression techniques are given in \cite{Smith:2015}.

While construction of the WCDs at the HAWC observatory ended in early 2015,
currently 250 out of 300 water Cherenkov tanks are in data acquisition.
Therefore, we will use the 250 tank configuration of the detector (HAWC-250) as
the baseline for this study.

\section{Sensitivity of HAWC to the Flux of $e^- + e^+$}

\begin{figure}[ht]
  \includegraphics[width=\textwidth]{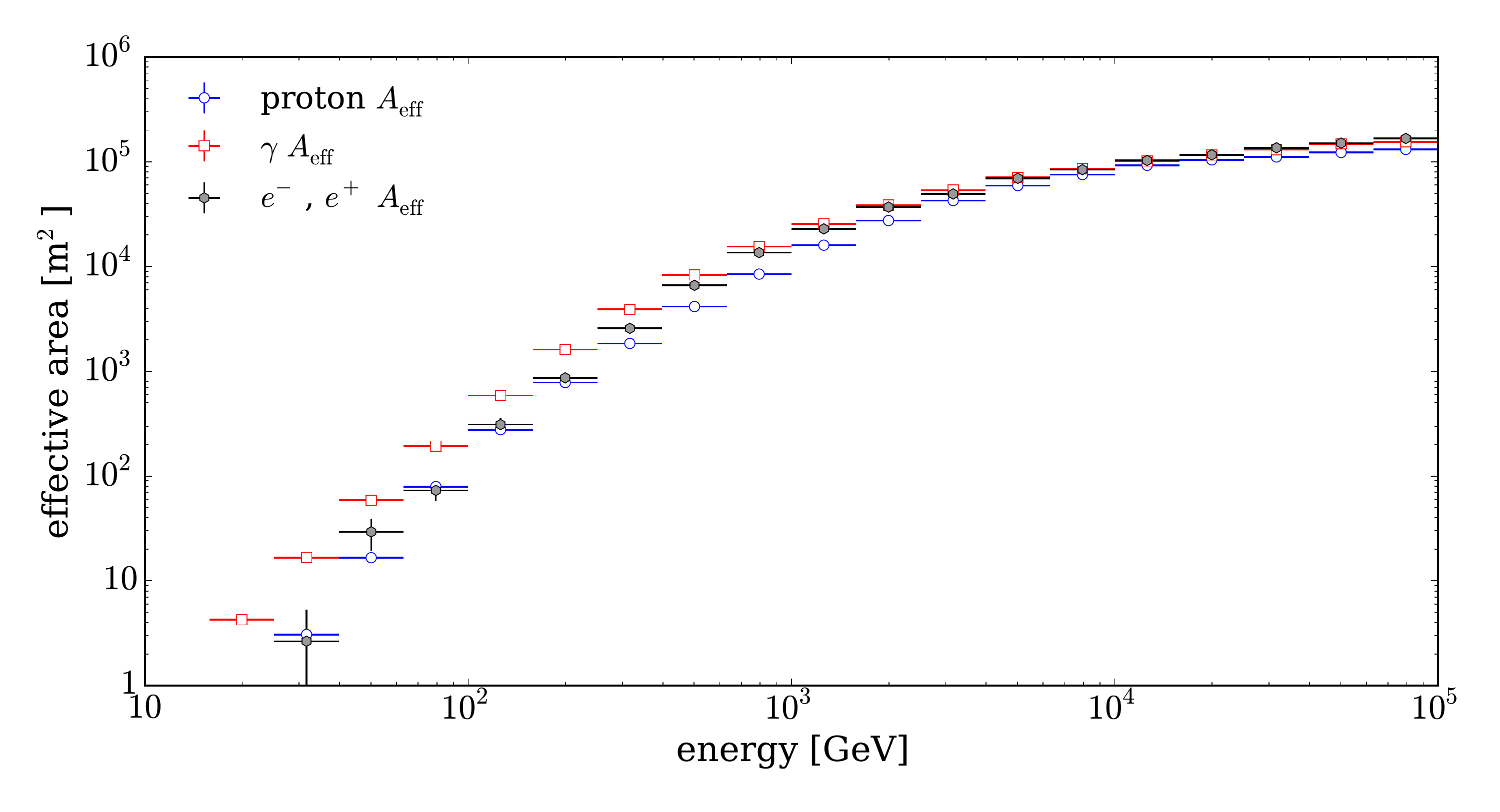}
  \caption{\label{fig:effective_area}} Simulated effective area of the HAWC-250
  observatory after the application of basic quality cuts to the event
  reconstruction (see text for details).  Air showers produced by protons are
  plotted as open circles, gamma rays as open squares, and electrons and
  positrons as filled hexagons.
\end{figure}

With an instantaneous field of view of about 2~sr, an uptime $>90\%$, and the
capability to discriminate electromagnetic air showers from hadronic air
showers, HAWC can be used to observe the isotropic \electron and \positron
flux.  The simulated effective area of HAWC for protons, gamma rays, and
electrons is plotted as a function of energy in Fig.~\ref{fig:effective_area}.
Air shower particles were produced using CORSIKA \cite{Heck:1998vt} and the
HAWC-250 detector response was simulated with GEANT4 \cite{Agostinelli:2002hh}.
The effective areas plotted in Fig.~\ref{fig:effective_area} were produced by
applying a simple PMT multiplicity cut of $N_\text{hit}\geq27$ (similar to
running conditions), a zenith angle cut of $\theta\leq50^\circ$, and by cutting
showers reconstructed $>2.5^\circ$ from their true direction. No further
quality cuts were applied.

Above several hundred GeV, where the HAWC array approaches high trigger
efficiency, the effective area for showers initiated by cosmic electrons is
similar to the effective areas for protons and $\gamma$-rays. As a result, the
event rate of electron-induced air showers is expected to be roughly $15$~Hz
in the full detector.  This implies $\sim5\times10^8$ \electron and \positron
showers observed per year with HAWC.

Unfortunately this isotropic flux is affected by multiple sources of background
which need to be modeled or removed. The background includes:
\begin{itemize}
  \item A much larger, but in principle reducible, isotropic flux of hadronic
        cosmic rays.
  \item A reducible flux of gamma rays from point sources and Galactic diffuse
        emission.
  \item An irreducible background of isotropic extragalactic gamma rays.
\end{itemize}
The flux of hadronic cosmic rays is the most significant source of background
for this analysis.  Because both the signal and background are isotropic, and
the signal spectrum is softer than the background spectrum, the only way to
measure the \eeflux flux is with very strong suppression of the hadronic cosmic
rays.  The relevant figure of merit for this suppression is the $Q$-value of
the analysis, i.e., the ratio of the signal selection efficiency
$\epsilon_\gamma$ to the background efficiency $\epsilon_\text{CR}$,
\[
  Q=\epsilon_\gamma/\sqrt{\epsilon_\text{CR}}.
\]
The current analysis, optimized for point sources of $\gamma$-rays, has
achieved values of $Q\approx5$ above several TeV \cite{Abeysekara:2013tza}.
This is sufficient to detect point sources but not an isotropic flux.  By
comparison, a value of $Q\approx30$ would be needed to achieve a
signal/background ratio of $1:1$ in this analysis.  The HESS Collaboration was
able to achieve a background suppression factor of $10^4$ (implying
$Q\approx100$) for their measurement of the \eeflux flux
\cite{Aharonian:2008aa,Aharonian:2009ah}.

In contrast to the cosmic-ray backgrounds, the $\gamma$-ray backgrounds in this
analysis are less significant.  Point sources and diffuse emission from the
Galaxy can be avoided by masking out those regions of the sky. The remaining
background from extragalactic isotropic $\gamma$-rays cannot be masked out and
will pass the selection cuts for electromagnetic showers. However, the
extragalactic flux is not expected to contribute significantly to the isotropic
signal below $10$~TeV.  In Fig.~\ref{fig:electron_positron_flux} we plot a
conservative estimate of the contamination of the $e^+e^-$ signal due to the
isotropic $\gamma$-ray background, extrapolating the $E^{-2.41}$ flux published
by Fermi-LAT in 2010 \cite{Abdo:2010nz}. But the actual contamination is
expected to be low in the region of interest ($<10$~TeV) since more recent
measurements of the isotropic $\gamma$ rays indicate a spectral cutoff at
$250$~GeV \cite{Ackermann:2014usa}.

The current state of the analysis suggests that due to cosmic-ray backgrounds,
it will be difficult to observe the isotropic \eeflux spectrum with HAWC.
However, the background suppression technique presented in
\cite{Abeysekara:2013tza} is admittedly na\"{i}ve. Recent efforts to improve
the cosmic-ray rejection power of the analysis indicate that improvements to
$Q=10$ are straightforward \cite{Hampel:2015,Capistran:2015}, and $Q\approx30$
is possible. These improvements will be the focus of future work.

\section{Sensitivity to the Moon Shadow}

An extension of the measurement of the \positron fraction in the cosmic ray
flux to TeV is well-motivated. There are currently no measurements above
$1$~TeV, and while the AMS Collaboration has reported evidence for a turnover
in the positron fraction at several hundred GeV (see \cite{Aguilar:2013qda} and
Fig.~\ref{fig:electron_positron_flux}) this feature is not sufficient to
distinguish between models of \positron acceleration and dark matter
annihilation.

HAWC is capable of discriminating \positron and \electron showers by observing
the shadow of the Moon in the \eeflux flux.  The Moon shadow is the small
deficit in the isotropic flux created by the absorption of electrons and
positrons in the lunar surface.  However, electrons and positrons from the
position of the Moon are deflected in equal but opposite directions by the
geomagnetic field. Hence, if the flux of positrons is sufficient, deflections
in the geomagnetic field should produce two observable deficits: a shadow in
the \electron flux, and a second displaced shadow in the \positron flux.

The average deflection of a particle of charge $Z$ and
energy $E$ at the location of HAWC is \cite{Abeysekara:2013qka}
\begin{equation}
  \delta\theta\approx 1.6^\circ\cdot Z\left(\frac{E}{\text{TeV}}\right)^{-1}.
\end{equation}
The median energy of cosmic rays observed in HAWC is about $2$~TeV. At this
energy the geomagnetic deflection will be $\sim0.8^\circ$, larger than both the
angular diameter of the Moon and the point spread function of the detector
($<0.5^\circ$ above $1$~TeV).

The use of the geomagnetic field to discriminate \positron from \electron is
similar to the analysis carried out with Fermi-LAT \cite{FermiLAT:2011ab} and
the proposal to observe the Moon shadow with the MAGIC telescope
\cite{Colin:2011wc}. Unlike MAGIC, HAWC is not affected by moonlight or weather
conditions and can be used to observe the Moon during all times and seasons.
However, the detector has considerably less background suppression capability
than an imaging air Cherenkov telescope, so we expect the sensitivity of the
two techniques to be approximately the same.

\begin{figure}[ht]
  \includegraphics[width=\textwidth]{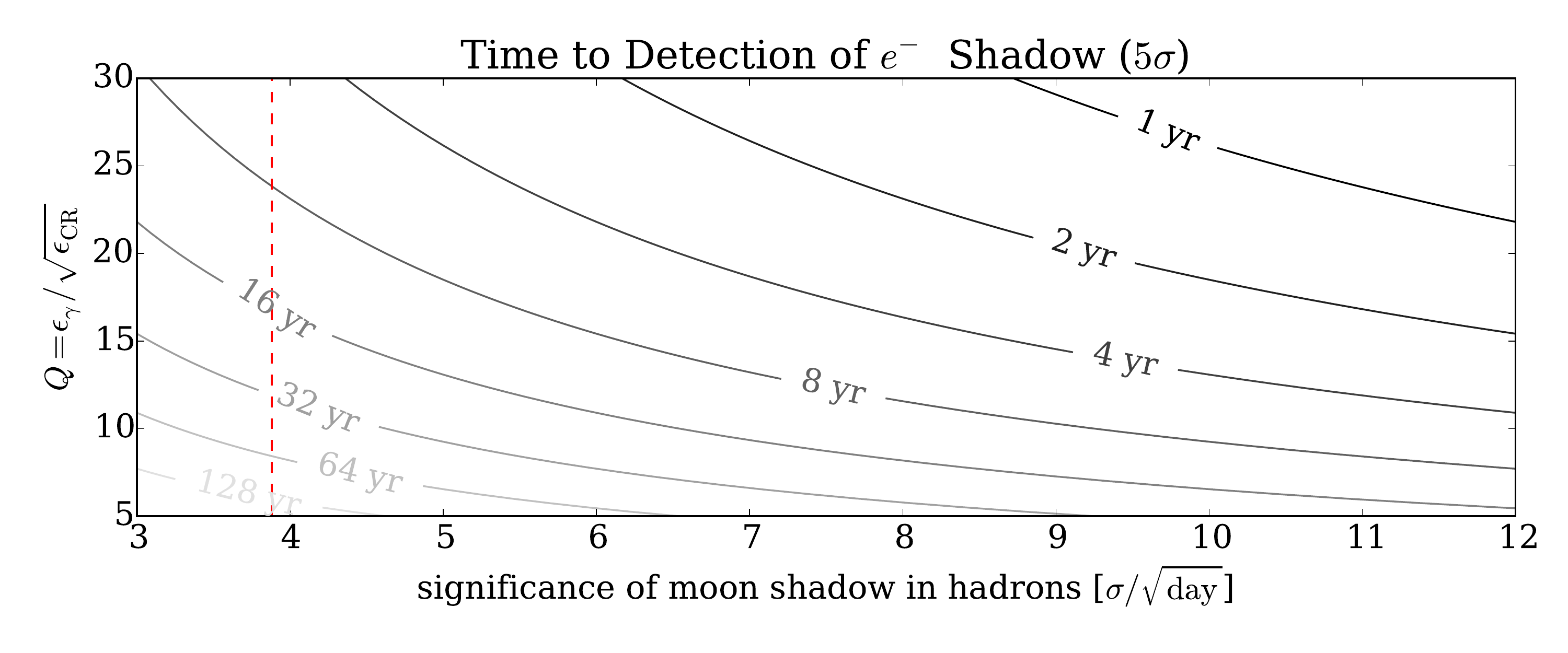}
  \caption{\label{fig:electron_shadow} Expected time to detect the
  electron Moon shadow at the $5\sigma$ level as a function of the
  daily significance of the shadow in hadronic cosmic rays and the
  $Q$-value of the analysis cuts. The current daily significance of the Moon
  shadow in cosmic rays is $<4\sigma$ and is indicated by a dashed line.}
\end{figure}

To estimate the time needed to observe the \electron shadow at the $5\sigma$
level, we compute the number of transits needed to observe the shadow given the
known sensitivity of the detector to the observed shadow in hadronic cosmic
rays:
\begin{equation}
  N = \left(\frac{5}{\sigma_\text{CR}^{N=1}}\cdot\frac{1}{Q}\cdot\frac{\Phi_\text{CR}}{\Phi_{e^-}}\right)^2.
\end{equation}
In this expression, $\sigma_\text{CR}^{N=1}$ represents the significance of
the observation of the Moon shadow in cosmic rays in one transit, and
$\Phi_\text{CR}/\Phi_{e^-}$ is the ratio of the fluxes of hadrons and
electrons.

Fixing the flux ratio and varying $\sigma_\text{CR}^{N=1}$ and $Q$, we produce
the plot shown in Fig.~\ref{fig:electron_shadow}. In this figure, we note that
there are many reasonable combinations of $\sigma_\text{CR}^{N=1}$ and $Q$
that define a path to the detection of the Moon shadow.  For example, if
$\sigma_\text{CR}^{N=1}$ is between $5\sigma$ and $10\sigma$ and $Q$ is
between $15$ and $25$, the $e^-$ shadow will be visible with $2$ to $4$ years
of data.

The current significance of the cosmic-ray Moon shadow in data is $\sim4\sigma$
per transit \cite{Fiorino:2015}. This is lower than the prediction from
simulations, mainly because the observed detector point spread function is
larger than expected.  We expect a detection of $>5\sigma$ per transit will be
achieved as our reconstruction techniques and understanding of the detector
improve.  In addition, the point spread function for the reconstruction of
\electron and \positron showers should be better than for hadronic cosmic rays
because the shower axis fit is optimized for electromagnetic showers.

The reduction of $\epsilon_\text{CR}$ and corresponding improvement in $Q$ is
also critical to this analysis.  Fig.~\ref{fig:electron_shadow} indicates that
$Q>15$ (corresponding $\epsilon_\text{CR}<0.5\%$) is likely necessary to
observe the \electron shadow during the lifetime of the experiment.
Simulations indicate that this level of background suppression and higher is
possible with more sophisticated gamma-hadron discrimination than what is
currently used in the $\gamma$-ray analysis \cite{Hampel:2015,Capistran:2015}.

\section{Conclusion}

HAWC is a high-uptime air shower array that observes $2/3$ of the sky the each
day.  While the detector was designed to search for pointlike and extended
sources of $\gamma$-rays on top of the large background of isotropic cosmic
rays, its gamma-hadron discrimination capability can also be used to observe
the flux of cosmic electrons and positrons.

As of this writing there are only two published measurements of the \eeflux
flux at TeV.  This situation will change during the next few years as
experiments such as AMS, CALET \cite{Adriani:2015cda}, and ground-based
observatories improve their methods and statistics.  HAWC can contribute to
these measurements, though a direct measurement of the leptonic flux is a
challenge because of its soft spectrum and the need for very strong hadronic
shower suppression.  Simulations indicate that the background rejection power
of HAWC can be improved substantially and these improvements will be the focus
of future work.

HAWC can also be used to estimate the positron fraction above $1$~TeV because
the geomagnetic field should create two well-separated deficits, or shadows, in
the \positron and \electron flux due to absorption of cosmic rays by the Moon.
Given expected improvements in the angular resolution of the detector and
optimistic but achievable improvements in the background reduction, the
\electron shadow could be detectable at $5\sigma$ within $2$ to $4$ years.

\section*{Acknowledgments}
\footnotesize{
We acknowledge the support from: the US National Science Foundation (NSF);
the US Department of Energy Office of High-Energy Physics;
the Laboratory Directed Research and Development (LDRD) program of
Los Alamos National Laboratory; Consejo Nacional de Ciencia y Tecnolog\'{\i}a
(CONACyT),
Mexico (grants 260378, 55155, 105666, 122331, 132197, 167281, 167733);
Red de F\'{\i}sica de Altas Energ\'{\i}as, Mexico;
DGAPA-UNAM (grants IG100414-3, IN108713,  IN121309, IN115409, IN111315);
VIEP-BUAP (grant 161-EXC-2011);
the University of Wisconsin Alumni Research Foundation;
the Institute of Geophysics, Planetary Physics, and Signatures at Los Alamos
National Laboratory;
the Luc Binette Foundation UNAM Postdoctoral Fellowship program.
}

\bibliography{icrc2015-0216}

\end{document}